\documentclass[12pt,preprint]{aastex}

\slugcomment{}

\shorttitle{Large-scale filamentary structure}
\shortauthors{Matsuda et al.}

\begin{document}


\title{Large-scale Filamentary Structure around the Protocluster at Redshift $z=3.1$\altaffilmark{1}}


\author{Yuichi Matsuda      \altaffilmark{2},
        Toru Yamada         \altaffilmark{3},
        Tomoki Hayashino    \altaffilmark{4},
        Hajime Tamura       \altaffilmark{4},
        Ryosuke Yamauchi    \altaffilmark{4},
        Takashi Murayama    \altaffilmark{5},
        Tohru Nagao         \altaffilmark{3,6},
        Kouji Ohta          \altaffilmark{2},
        Sadanori Okamura    \altaffilmark{7,8},
        Masami Ouchi        \altaffilmark{9},
        Kazuhiro Shimasaku  \altaffilmark{7,8},
        Yasuhiro Shioya     \altaffilmark{5}, and 
        Yoshiaki Taniguchi  \altaffilmark{5}
}

\email{matsdayi@kusastro.kyoto-u.ac.jp}

\altaffiltext{1}{Based on data collected at Subaru Telescope which is operated by the National Astronomical Observatory of Japan.}
\altaffiltext{2}{Department of Astronomy, Kyoto University, Sakyo-ku, Kyoto 606-8502, Japan}
\altaffiltext{3}{National Astronomical Observatory of Japan, Mitaka, Tokyo 181-8588, Japan}
\altaffiltext{4}{Research Center for Neutrino Science, Graduate School of Science, Tohoku University, Aramaki, Aoba, Sendai 980-8578, Japan}
\altaffiltext{5}{Astronomical Institute, Graduate School of Science, Tohoku University, Aramaki, Aoba, Sendai 980-8578, Japan}
\altaffiltext{6}{INAF -- Osservatorio Astrofisico di Arcetri, Largo Enrico Fermi 5, 50125 Firenze, Italy}
\altaffiltext{7}{Department of Astronomy, School of Science, University of Tokyo, Tokyo 113-0033, Japan; okamura@astron.s.u-tokyo.ac.jp}
\altaffiltext{8}{Research Center for the Early Universe, School of Science, University of Tokyo, Tokyo 113-0033, Japan}
\altaffiltext{9}{Space Telescope Science Institute, 3700 San Martin Drive, Baltimore, MD 21218, USA}


\begin{abstract}

 We report the discovery of a large-scale coherent filamentary structure 
of Ly$\alpha$ emitters in a redshift space at $z=3.1$. We carried out 
spectroscopic observations to map the three dimensional structure of the 
belt-like feature of the Ly$\alpha$ emitters discovered by our previous 
narrow-band imaging observations centered on the protocluster at $z=3.1$. 
The feature was found to consist of at least three physical filaments 
connecting with each other. The result is in qualitative agreement with 
the prediction of the 'biased' galaxy-formation theories that galaxies 
preferentially formed in large-scale filamentary or sheet-like mass 
overdensities in the early Universe. We also found that the two known 
giant Ly$\alpha$ emission-line nebulae showing high star-formation 
activities are located near the intersection of these filaments, which 
presumably evolves into a massive cluster of galaxies in the local Universe. 
This may suggest that massive galaxy formation occurs at the characteristic 
place in the surrounding large-scale structure at high redshift. 

\end{abstract}


\section{INTRODUCTION}

 Theories of structure formation predict that galaxy formation 
preferentially occurs along large-scale filamentary or sheet-like 
mass overdense regions in the early Universe and the intersections 
of such filaments or sheets evolve into dense clusters of galaxies 
at the later epoch (Governato et al. 1998; Kauffmann et al. 1999; 
Cen \& Ostriker 2000; Benson et al. 2001). Recent deep observations 
of star-forming galaxies at high redshift, such as Lyman break galaxies 
(LBGs) or Ly$\alpha$ emitters (LAEs), have revealed their inhomogeneous 
spatial distribution over several tens to a hundred Mpc (Steidel et al. 
1998, 2000, hereafter S98, S00; Adelberger et al. 1998; M\"{o}ller \& Fynbo 
2001, Shimasaku et al. 2003, 2004; Palunas et al. 2004; Francis et al. 2004; 
Ouchi et al. 2005). However, their volume coverage or redshift information 
are still limited. Thus, there is little direct observational evidence of 
this theoretical prediction.

 A large overdensity of LBGs and LAEs was discovered at $z=3.1$ in the 
SSA22 and it was regarded as a protocluster of galaxies (S98 and S00).
We carried out wide-field and deep narrow-band imaging observations 
of the SSA22 proto-cluster and discovered a belt-like region of high 
surface density of LAEs with the length of more than 60 Mpc and the 
width of about 10 Mpc in comoving scale (Hayashino et al. 2004, 
hereafter H04). We could not distinguish, however, whether the 
belt-like feature on the sky is a physically coherent structure or 
just a chance projection of isolated clumps in space.

 There exist two giant extended Ly$\alpha$ nebulae (Ly$\alpha$ blobs, LABs) 
in this proto-cluster whose sizes are larger than 100 kpc (S00). We 
also detected in our narrow-band images 33 LABs with Ly$\alpha$ isophotal 
area larger than 16 arcsec$^2$, which corresponds to 900 kpc$^2$ or 
${\rm d}\approx 30$ kpc at $z=3.1$ (Matsuda et al. 2004, hereafter M04). 
It is, however, noted that M04's LABs are smaller than S00's two giant 
LABs. The two giant LABs seem to be rare outstanding objects in the
region.

 In this letter, we present the redshift distribution of LAEs in 
the belt-like feature. We use AB magnitudes and adopt a set of 
cosmological parameters, $\Omega_{\rm M} = 0.3$, $\Omega_{\Lambda} 
= 0.7$ and $H_0 = 70$ km s$^{-1}$ Mpc$^{-1}$.

\section{OBSERVATIONS}

 The targets of our spectroscopy were chosen from the 283 candidate 
LAEs detected in our narrow-band imaging observations centered 
on the SSA22 proto-cluster (RA=$22^{\rm h}17^{\rm m}34^{\rm s}.0$, 
Dec=$+00^{\rm o}17'01''$, [J2000.0]) at $z=3.1$ using the Prime Focus 
Camera (Suprime-Cam, Miyazaki et al. 2002) of the 8.2 m Subaru telescope. 
We briefly describe the selection criteria below. Details are given in H04. 
The 283 candidate LAEs satisfies the following criteria; 
(1) bright narrow-band magnitude at 4970\AA , $NB497<25.8$ mag, and 
(2) large observed equivalent width, $EW_{\rm obs}>154$ \AA\ 
(or L$({\rm Ly}\alpha)>10^{41}$ erg s$^{-1}$ at $z=3.1$). In addition, 
we imposed one of the following two criteria; 
(3) red continuum color, $B-V_c>0.2$ mag, for the objects with $V_c$ 
brighter than 26.4 mag, or (4) larger equivalent width, 
$EW_{\rm obs} > 267$ \AA, for the objects with $V_c$ fainter than 26.4 
mag. Here, $V_c$ represents the emission-line free $V$-band magnitude 
obtained after subtracting the narrow-band flux from the $V$-band flux. 
Note that LAEs at $z \simeq 3$, which often have continuum spectra of 
$f_{\nu}\sim\nu^0$, should have $B-V_c$ greater than 0.2 mag, since 
the Ly$\alpha$ forest dims the continuum blueward of Ly$\alpha$ line 
of galaxies (e.g. Madau 1995). 

 The distribution of the 283 LAEs showed the belt-like feature 
of high surface density running from NW to SE on the sky (the 
bottom right panel of Figure 1). In order to examine the 
three-dimensional structure of this belt-like feature, we carried 
out the spectroscopic observations of the candidate LAEs using 
Faint Object Camera And Spectrograph (FOCAS, Kashikawa et al. 2002) 
on the Subaru telescope in the two first-half nights of 29 and 30 
October 2003 (UT). The diameter of the field of view of a FOCAS slit
mask was $6'$. We used 6 slit masks along the belt-like feature
(see Figure 1). Out of the 283 LAEs, 122 are located in the area
covered by the 6 masks. We prepared the slits for bright 84 of the
122 selected in order of NB magnitudes. We used the 300B grism and 
the SY47 filter. The typical wavelength coverage was 4700--9400 \AA . 
The slit widths were $0''.8$ and the seeing was $0''.4$--$1''.0$. 
The spectral resolution was ${\rm FWHM} \sim 10$ \AA\ at 5000 \AA , 
corresponding to a velocity resolution of $\sim$600 km s$^{-1}$. 
The pixel sampling was 1.4 \AA\ in wavelength (no binning) and $0''.3$
in spatial direction (3 pixel binning). Exposure times were 0.7--1.5 
hours per mask. We reduced the data using the IDL program developed by 
FOCAS team and IRAF. We used both Th-Ar lamp spectra and night sky 
lines in wavelength calibration. The rms of fitting error in wavelength 
calibration was smaller than 2 \AA . We used the standard star Feige 110 
in flux calibration.

\section{RESULTS AND DISCUSSION}

 Among the 84 targeted LAEs, 56 spectra show a single emission-line 
with a signal-to-noise ratio (S/N) per resolution element (10 \AA ) 
larger than 5. The typical S/N of the 56 emission-lines is about 11. 
The most plausible interlopers of the emission-lines are 
[OII]$\lambda$3727 at $z=0.325-0.346$. In this case, 
the [OIII]$\lambda\lambda$4959,5007 should be detected, since 
the wavelength coverage of most spectra extends longer 
than 6700\AA. Most of star-forming galaxies at low redshifts 
have [OIII]$\lambda$5007/[OII]$\lambda$3727 ratio larger than 
0.15 (e.g. Jansen et al. 2000). Since the upper limits for 
[OIII]$\lambda$5007/[OII]$\lambda$3727 ratios are smaller 
than 0.15 for our spectra, there is no evidence for contamination 
of [OII] emission-line galaxies. Therefore, we identify the single 
emission-lines as Ly$\alpha$ with high confidence. The mean redshift 
of the 56 identified LAEs is $<z>=3.091$ and the redshift dispersion 
is $\Delta{z}=0.015$ ($\Delta{v}=1100$ km/s). We stacked the observed 
frame spectra of remaining 28 unidentified objects with emission-lines 
with S/N$<5$. The stacked spectrum shows a significant double-peak 
emission-line, whose profile is similar to the shape of the redshift 
histogram of the 56 identified LAEs. Accordingly, it is highly probable 
that a large fraction of the unidentified LAEs is also located in the 
same structure at $z\sim3.1$.

 Redshifts of the LAEs are not the exact measure of the Hubble flow or 
the spatial distribution due to their peculiar velocities. In fact, 
peculiar velocities are considered to be the dominant source of errors 
in the spatial distribution of LAEs\footnote{We note that there are 
differences between the redshifts of Ly$\alpha$ emission-lines and 
those of other nebular lines which are expected to be better tracers 
of the systemic redshift of galaxies (Adelberger et al. 2003). Since 
neutral hydrogen and dust in galaxies tend to absorb the blue side of 
the Ly$\alpha$ emission-lines, the peak of Ly$\alpha$ emission-lines 
apparently shifts to higher redshifts. According to Adelberger et al. 
(2003), the excess and the rms scatter of redshifts for Ly$\alpha$ 
emission-line for LAEs are $310 ~{\rm km}~{\rm s}^{-1}$ and 
$250 ~{\rm km}~{\rm s}^{-1}$, respectively. This scatter is smaller 
than the predicted peculiar velocity dispersion.}. While the peculiar 
velocity dispersion of galaxies is $500 - 600 ~{\rm km}~{\rm s}^{-1}$ in 
the local universe (Zehavi et al. 2002; Hawkins et al. 2003), it is expected 
to be smaller at high redshifts even in the over-dense regions (Hamana 
et al. 2001, 2003). Indeed, the predicted pair wise peculiar velocity 
dispersion of galaxies at $z\sim 3$ in cosmological simulations is 
$300 - 400 ~{\rm km}~{\rm s}^{-1}$ (Zhao et al. 2002), which corresponds 
to a very small redshift dispersion of $\sigma_{z} \sim 0.005$.

 In Figure 1, we plot the resultant three dimensional distribution 
of the 56 LAEs, together with the projected distribution. We can see 
that the belt-like feature seen in the projected distribution consists 
of a filamentary structure running from ($\Delta$RA[arcmin], 
$\Delta$Dec[arcmin], Redshift[$z$])$\approx$(25, 18, 3.108) to 
(5, 8, 3.088) and another concentration around (19, 14, 3.074). 
In order to compute the volume density of the LAEs, we convolved the 
spatial distribution of the 56 LAEs with a Gaussian kernel with 
$\sigma=4$ Mpc, which is comparable to the redshift dispersion due 
to the predicted peculiar velocity dispersions, 
$\Delta v\sim 400$ km s$^{-1}$. By drawing the projected contour of 
volume density of $2.0 \times 10^{-3}~{\rm Mpc}^{-3}$, we identified 
three filaments connecting with each other with the intersection at 
around (16, 11, 3.094). The length of each filament is about 30 Mpc 
and the width is about 10 Mpc in comoving scale. This is the largest 
coherent filamentary structure mapped in three dimensional space at 
$z \ge3 $. In the central $8.7' \times 8.9'$ region, we also plot in 
Figure 1 the 21 LBGs in Steidel et al. (2003) whose Ly$\alpha$ emission 
redshifts lie in the same redshift range of our narrow-band imaging 
observations, $z=3.054-3.120$. These LBGs seem to be concentrated near 
the intersection of the filaments of LAEs.

 Although our spectroscopic observations are not complete yet, 
we tried to constrain the three-dimensional over-density of these 
filaments using the same Gaussian kernel above. The highest number 
density of the LAEs is $\approx 6.0\pm 2.4 \times 10^{-3}~{\rm Mpc}^{-3}$. 
The average number density along the filaments is 
$\approx 3.0 \times 10^{-3}~{\rm Mpc}^{-3}$ while the average
number density of 283 LAEs in the whole volume sampled by the 
narrow band filter, $1.4 \times 10^5$ Mpc$^3$, is 
$2.0 \times 10^{-3}~{\rm Mpc}^{-3}$. Note, however, that the number 
density estimated from the spectroscopy is the lower limit because 
of the incompleteness in our redshift measurements. If we assume that 
the remaining 66 LAEs, which are in the fields of 6 slit masks but not 
considered in the present analysis, have similar spatial distribution, 
the real number density of the LAEs in the filament would be higher 
by a factor of 2 and would be 3 times as large as the average value of 
the entire field.

 In Figure 2, we show the composite spectrum of the 56 LAEs which 
are shifted into the rest frame using their Ly$\alpha$ emission 
redshifts. There is no evidence of [OIII] emission-lines for [OII] 
emitters at $z\approx 0.33$ in this spectrum. The rest frame 
EW of Ly$\alpha$ emission-line of the spectrum is about 60 \AA , 
which is roughly consistent with the EW estimated from our narrow-band 
imaging observations. The continuum flux blueward of the Ly$\alpha$ 
line is $\sim$20\% dimmer than that redward of the line. However, 
this value is small compared with the continuum depression found in 
high S/N composite spectra of LBGs (e.g. Shapley et al. 2003). This 
may be due to the poor quality of our spectra blueward of Ly$\alpha$ 
emission-line because of the rapid decrease of the transmittance of 
our SY47 filter at 4600--4900\AA . The spectral resolution of 
$600 ~{\rm km}~{\rm s}^{-1}$ is too large to resolve the line 
profile of Ly$\alpha$, since the typical FWHM of the Ly$\alpha$ 
emission line of LAEs at high redshifts is $300 ~{\rm km}~{\rm s}^{-1}$ 
(e.g. Venemans et al. 2005). We could not find any significant evidence 
of other emission or absorption lines in the composite spectrum. 
The observed CIV$\lambda$1549/Ly$\alpha$ ratio is smaller than 0.1. 
This suggests that narrow-line AGNs do not dominate these LAEs; 
the CIV$\lambda$1549/Ly$\alpha$ ratio is typically 0.2 for narrow-line 
AGNs at $z\sim 3$ (Steidel et al. 2002). Thus the Ly$\alpha$ emission 
of the LAEs is likely to originate from their star-formation activities.

 A cosmological numerical simulation of galaxy formation including gas 
particles suggests that the average star-formation rate (SFR) of 
galaxies at $z=3$ is nearly independent of their environment while 
a few galaxies with very large SFR may exist in the highest density 
regions (Kere\u{s} et al. 2005). We estimated the SFR of the LAEs from 
their UV continuum luminosities. The left panel of Figure 3 shows the 
SFR of all the 283 LAEs as a function of projected surface number density, 
while the right panel shows the SFR as a function of volume density
for the 56 LAEs with spectroscopic redshifts. We do not find any significant 
correlation. We checked that the dust reddening correction using their 
continuum colors does not change this trend. The environmental effect 
thus seems to be weak for the high-redshift LAEs, which is in fact 
consistent with the prediction of the simulation in Kere\u{s} et al. 
(2005).

 There are lines of evidence that the LAB1 is a very massive galaxy with 
intensive star-formation activities. Bright submillimeter source was 
detected at the position of LAB1 and the SFR estimated from the 
submillimeter flux is extremely large ($\sim 1000$ M$_{\odot}$ yr$^{-1}$)
 (Chapman et al. 2001). Not only the dust emission but the CO emission 
line was detected in LAB1, which implies that the large amount of molecular 
gas also exists (Chapman et al. 2004). The large mass of the host dark 
matter halo ($\sim10^{13}$ M$_{\odot}$) was also suggested for LAB1 from 
the velocity dispersion and the physical extent of Ly$\alpha$ emission-line, 
by assuming that the gas is bound within its gravitational potential 
(Bower et al. 2004). LAB2 looks similar to LAB1 on the Ly$\alpha$
image, and it is likely that the two giant LABs are very massive galaxies 
in their forming phase. Accordingly, it is interesting to see their 
location with respect to the filamentary structure.

 We measured the redshifts of LAB1 and LAB2 at their surface 
brightness peaks of the emission line. Their redshifts, $z=3.102$ for 
LAB1 and $z=3.103$ for LAB2, indicate that they are located near 
the intersection of the three filaments (Figure 1). Cosmological 
simulations predict that the intersections of large-scale filaments 
in the early Universe evolve into the present day massive clusters 
of galaxies. Thus, we can reasonably speculate that the two LABs may 
be progenitors of very massive galaxies near the center of a massive 
cluster. The smaller LABs of M04 are also concentrated near the position 
of the intersection in the projected distribution. It would be
interesting to investigate by future observations whether or 
not the smaller LABs are preferentially located at the intersection 
of filaments in three dimensional space.

\acknowledgments

 We thank the anonymous referee for useful comments which have 
significantly improved the paper. We thank the staff of the Subaru 
Telescope for their assistance with our observations. The research 
of T.Y. is partially supported by the grants-in-aid for scientific research 
of the Ministry of Education, Culture, Sports, Science, and Technology 
(14540234 and 17540224).

\clearpage



\begin{figure}
\epsscale{1.0}
\plotone{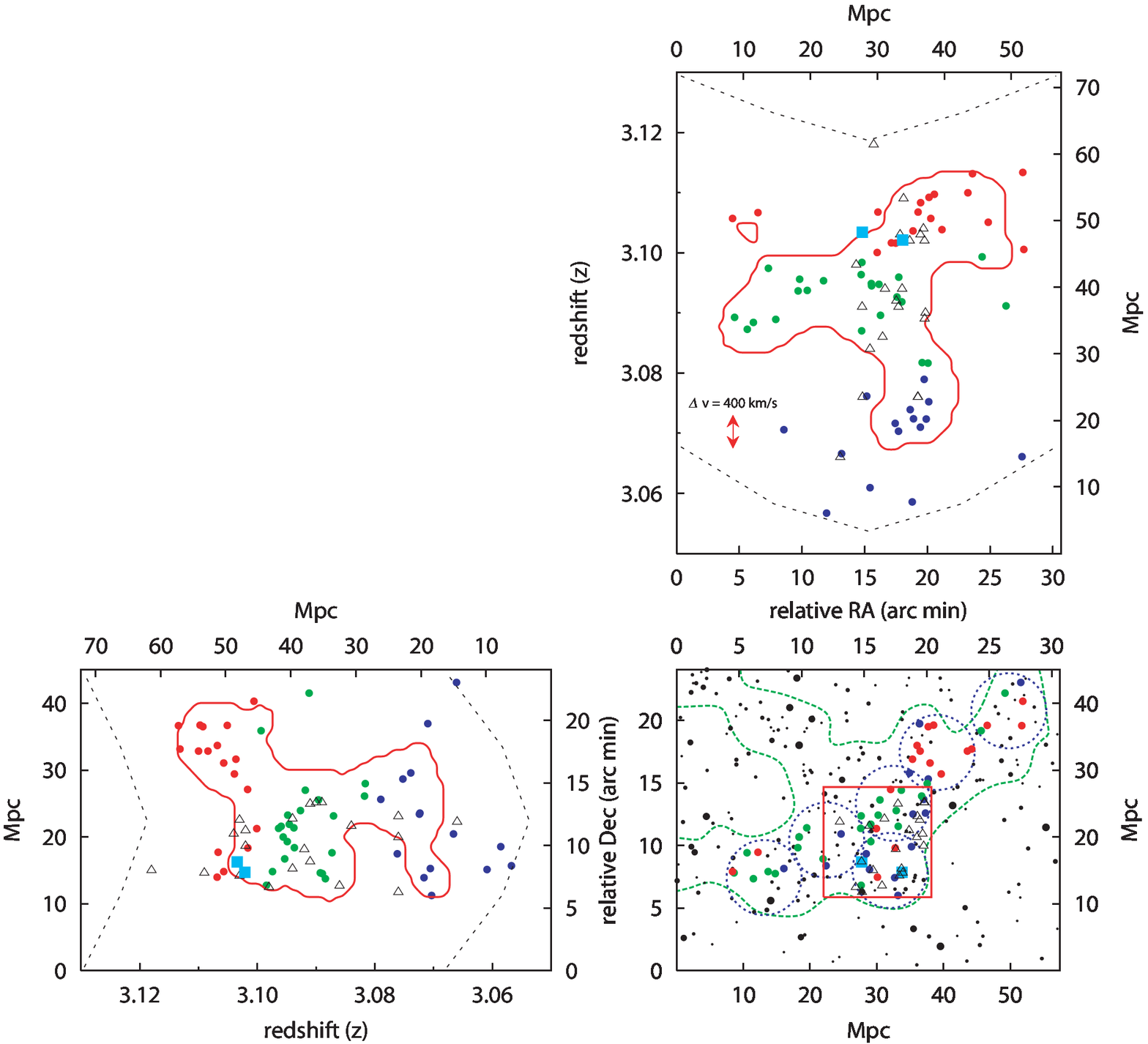}
\caption{({\it Bottom right panel}) 
The sky map of the 283 candidate LAEs detected in Hayashino
et al. (2004). The green line shows the average local surface 
density of LAEs in this field  (see the text).
Cyan squares show two giant LABs. Blue circles 
show the field of view of 6 masks. Blue, green, and red points 
show the LAEs at $z=3.05-3.08$, $z=3.08-3.10$, and at $z=3.10-3.12$, 
respectively. The triangles show the LBGs in the SSA22a field (the red box, 
$8.7' \times 8.9'$, Steidel et al. 2003). ({\it Top right and bottom 
left panels}) The redshift space distribution of 56 LAEs with spectroscopic 
redshifts. The red line shows the projected contour of the local volume 
density of LAEs of $2 \times 10^{-3}~{\rm Mpc}^{-3}$ (see the text). 
The predicted peculiar velocity dispersion of 400 km s$^{-1}$ is 
shown by red arrows. The dotted lines show the redshift range sampled 
within $\ge 50 \%$ of the peak transmittance of our narrow-band filter.
\label{fig1}}
\end{figure}

\clearpage
\begin{figure}
\epsscale{1.0}
\plotone{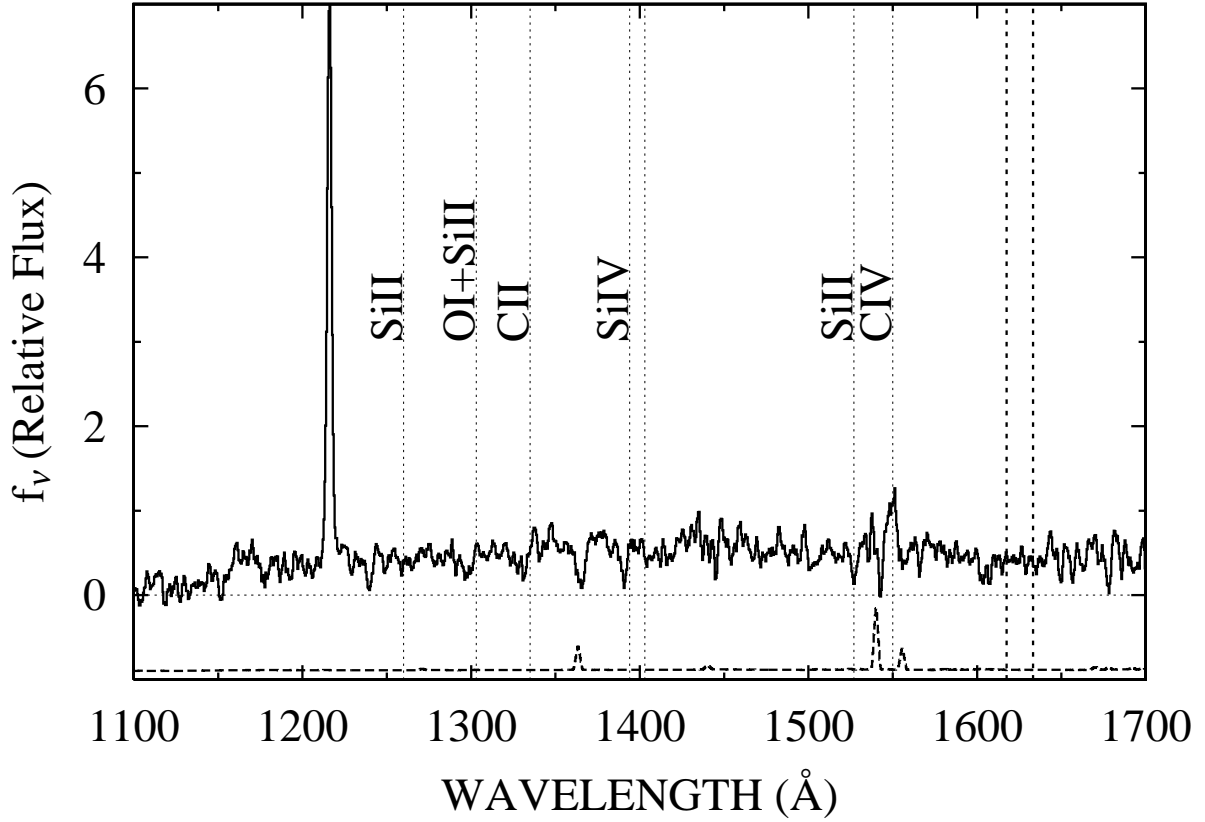}
\caption{The median composite spectrum of the 56 LAEs shown in the 
rest frame. The vertical short dashed lines show 
[OIII]$\lambda\lambda$4959,5007 for [OII]$\lambda$3727 emitters at 
$z=0.33$. The vertical dotted lines show wavelengths of strong 
interstellar absorption lines in Lyman break galaxies. The spectrum 
have been smoothed by a boxcar kernel of width 2.5\AA\ (the spectral 
resolution in the rest frame). The long dashed line shows the typical 
sky spectrum for LAEs at $z=3.09$.\label{fig2}}
\end{figure}

\clearpage
\begin{figure}
\epsscale{1.0}
\plottwo{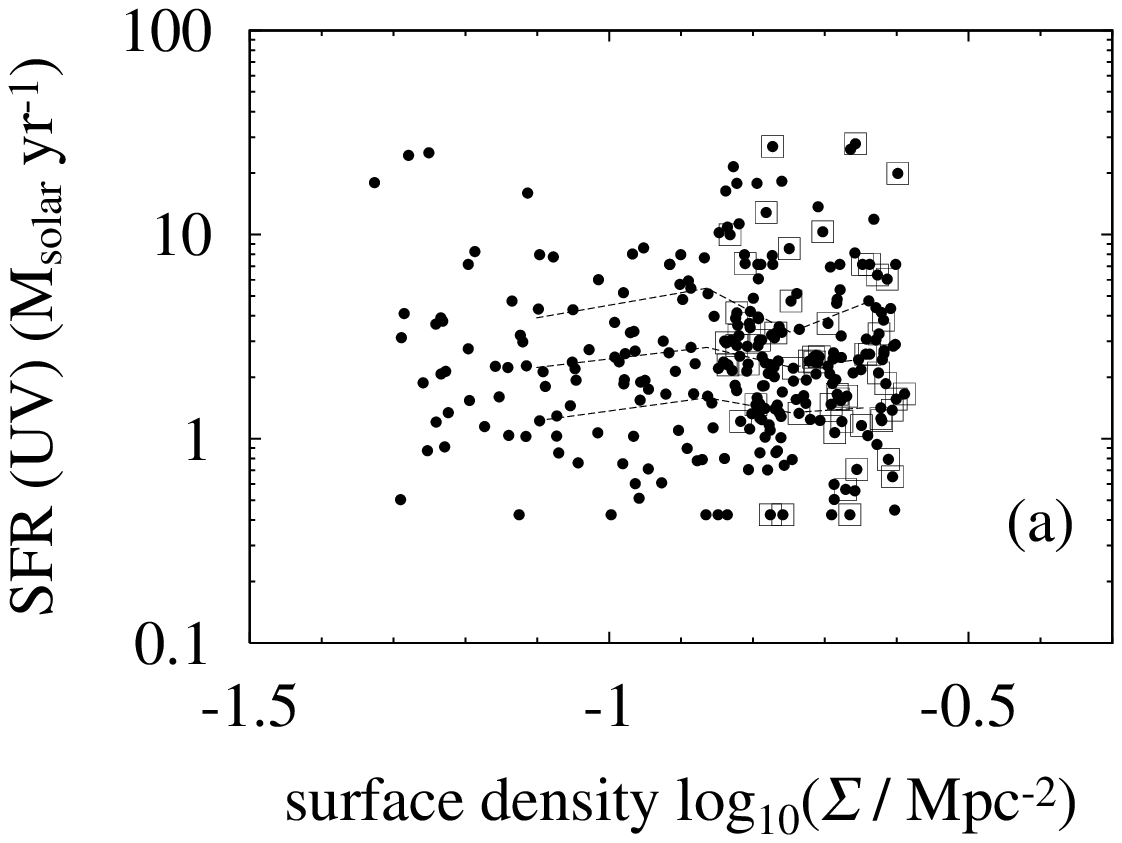}{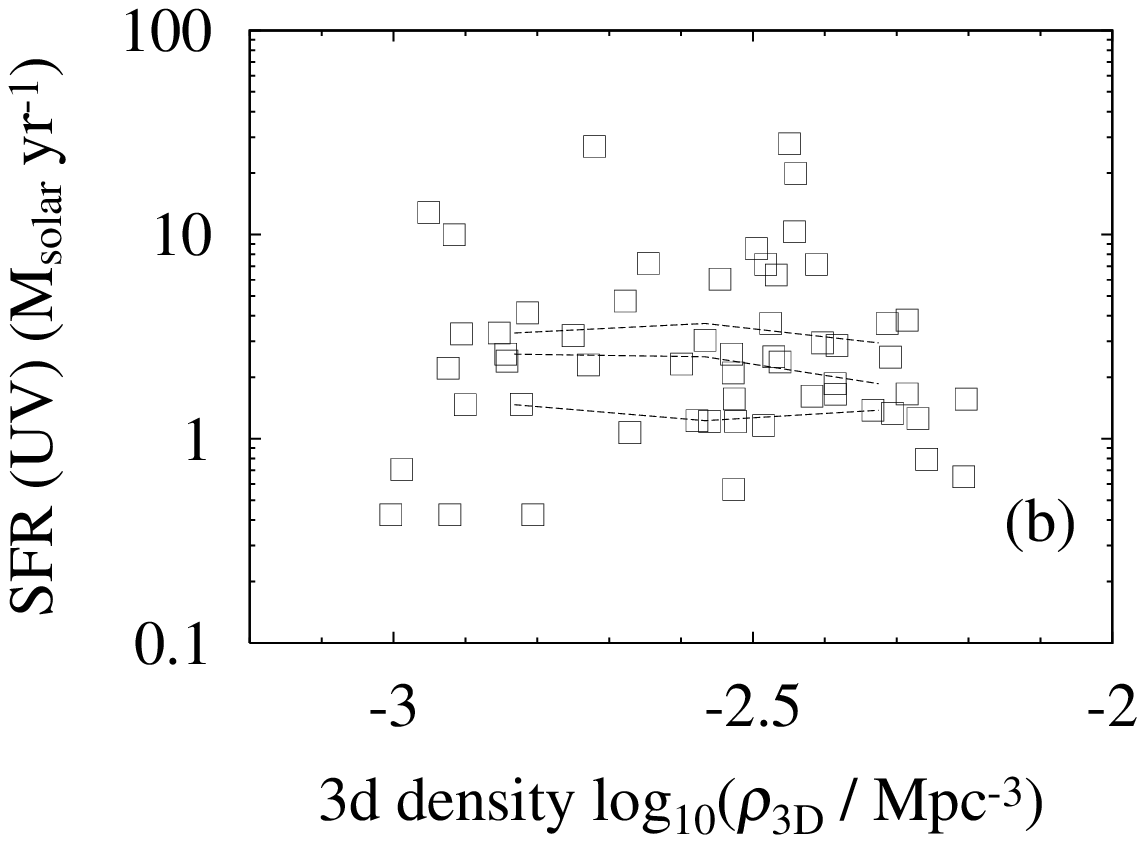}
\caption{The distribution of SFR estimated from UV continuum 
flux as a function of the projected local surface density 
$\Sigma$ of LAEs ((a),{\it left panel}) and three-dimensional 
density $\rho_{\rm 3D}$((b),{\it right panel}). The squares show 
LAEs with spectroscopic redshifts and the points show LAEs without 
spectroscopic redshifts. The three dashed lines show the 75th 
percentile, the median, and the 25th percentile from the top to 
bottom. The three-dimensional density $\rho_{\rm 3D}$ is not 
corrected for the incompleteness in our redshift measurement.
\label{fig3}}
\end{figure}









\clearpage








\clearpage

\begin{thebibliography}{}

\bibitem[{Adelberger et al. 1998}]{ad98} Adelberger, K. L. et al. 
1998, \apj, 505, 18

\bibitem[{Adelberger et al. 2003}]{ad03} Adelberger, K. L., 
Steidel, C. C., Shapley, A. E. \& Pettini, M. 2003, \apj, 584, 45

\bibitem[{Benson et al. 2001}]{be01} Benson, A. J., Frenk, C. S., 
Baugh, C. M., {Cole}, S., \& {Lacey}, C. G. 2001, \mnras, 327, 1041

\bibitem[{Bower et al. 2004}]{bo04} Bower, R. G. et al. 2004, 
\mnras, 351, 63

\bibitem[{Cen \& Ostriker 2000}]{ce00} Cen, R., \& Ostriker, J. P. 
2000, \apj, 538, 83

\bibitem[{Chapman et al. 2001}]{ch01} Chapman, S. C. et al. 2001, 
\apj, 548, L17

\bibitem[{Chapman et al. 2004}]{ch04} Chapman, S. C. et al. 2004, 
\apj, 606, 85

\bibitem[{Francis et al. 2004}]{fr04} Francis, P. J., Palunas, P., 
Teplitz, H. I., Williger, G. M., Woodgate, B. E. 2004, \apj, 614, 75

\bibitem[{Governato et al. 1998}]{go98} Governato, F. et al. 1998, 
\nat, 392, 359

\bibitem[{Hamana et al. 2001}]{ham01} Hamana, T., Colombi, S., 
\& Suto, Y. 2001, \aap, 367, 18

\bibitem[{Hamana et al. 2003}]{ham03} Hamana, T., Kayo, I., 
Yoshida, N., Suto, Y., \& Jing, Y. P. 2003, \mnras, 343, 1312.

\bibitem[{Hayashino et al. 2004}]{ha04} Hayashino, T. et al. 2004, 
\aj, 128, 2073

\bibitem[{Hawkins et al. 2003}]{ha03} Hawkins, E. et al. 2003, 
\mnras, 346, 78

\bibitem[{Jansen, et al. 2000}]{ja00} Jansen, R. A., Fabricant, D., 
Franx, M., \& Caldwell, N. 2000, \apjs, 126, 331  

\bibitem[{Kashikawa et al. 2002}]{ka02} Kashikawa, N. et al. 2002, 
PASJ, 54, 819

\bibitem[{Kauffmann et al. 1999}]{ka99} Kauffmann, G., Colberg, J. M., 
Diaferio, A., \& White, S. D. 1999, \mnras, 307, 529

\bibitem[{Kere\u{s} et al. 2004}]{ke04} Kere\u{s}, D., Katz, N., 
Weinberg, D. H. \& Dav\'{e}, R. 2005, \mnras, 363, 2 

\bibitem[{Matsuda et al. 2004}] {ma04} Matsuda, Y. et al. 2004, 
\aj, 128, 569

\bibitem[{Miyazaki et al. 2002}]{mi02} Miyazaki, S. et al. 2002, 
PASJ, 54, 833

\bibitem[{Moller \& Fynbo 2001}]{mo01} M\"{o}ller, P., \& Fynbo, J. U. 
2001, \aap, 372, L57 

\bibitem[{Ouchi et al. 2005}]{ou05} Ouchi, M. et al. 2005, \apj, 620, 
L10

\bibitem[{Palunas et al. 2004}]{pa04} Palunas, P., Teplitz, H. I., 
Francis, P. J., Williger, G. M., Woodgate, B. E. 2004, \apj, 602, 545

\bibitem[{Shapley et al. 2003}]{sha03} Shapley, A. E., Steidel, C. C., 
Pettini, M. \& Adelberger, K. L. 2003, \apj, 568, 65

\bibitem[{Shimasaku et al. 2003}]{shi03} Shimasaku, K. et al. 2003, 
\apjl, 566, L111

\bibitem[{Shimasaku et al. 2004}]{shi04} Shimasaku, K. et al. 2004, 
\apjl, 605, L93

\bibitem[{Steidel et al. 1998}]{st98} Steidel, C.~C. et al. 1998, 
\apj, 492, 428

\bibitem[{Steidel et al. 2000}]{st00} Steidel, C.~C. et al. 2000, 
\apj, 552, 170

\bibitem[{Steidel et al. 2002}]{st02} Steidel, C. C. et al. 2002, 
\apj, 576, 653

\bibitem[{Steidel et al. 2003}]{st03} Steidel, C. C. et al. 2003, 
\apj, 592, 728.

\bibitem[{Venemans et al. 2005}]{ve05} Venemans, B. P. et al. 2005, 
\aap, 431, 739

\bibitem[{Zehavi et al. 2002}]{ze02} Zehavi, I. et al. 2002, \apj, 
571, 172

\bibitem[{Zhao et al. 2002}]{za02} Zhao, D., Jing, Y. P. 
\& B\"{o}rner, G. 2002, \apj, 581, 876

\end{thebibliography}
\end{document}